\documentclass[12pt,a4paper]{iopart}

\usepackage[english]{babel}
\usepackage{iopams}
\usepackage{ae} %full vectorgraphic output.
\usepackage{subfigure}
\usepackage{color}
\usepackage{ifthen} 
\usepackage{bm}

\usepackage[pdftex]{graphicx}
\usepackage[pdftex]{hyperref} 
\hypersetup{colorlinks=true,
  pdfstartview=FitV,
  linkcolor=blue,
  citecolor=blue,
  urlcolor=blue,
  pdfauthor={HORVATH Laszlo, horvath.laszlo@reak.bme.hu},
  pdfsubject={Horvath phases article manuscript},
  pdftitle={Horvath phases article manuscript}
  }
\DeclareGraphicsExtensions{.pdf}

\newcommand{\eqref}[1]{(\ref{#1})}

\newcommand{\dt}{\mathrm{d}t}

\graphicspath{{figures/}}

\begin{document}

\title{Reducing systematic errors in time-frequency resolved mode number analysis}
% \title[Horvath mode number paper - Revision number: \red{\textbf{r\svnrev}} - \svndate]{Reducing systematic errors in time-frequency resolved mode number analysis}
\author{L. Horv\'{a}th$^1$, P. Zs. Poloskei$^1$, G. Papp$^{2,3}$, M. Maraschek$^2$, \\
K.H. Schuhbeck$^2$, G. I. Pokol$^1$, the EUROfusion MST1 Team\footnote{See \url{http://www.euro-fusionscipub.org/mst1}} and the ASDEX Upgrade Team$^2$}

\address{${}^1${}Institute of Nuclear Techniques, Budapest University of Technology and Economics, Pf 91, H-1521 Budapest, Hungary\\
  ${}^2${}Max-Planck-Institute for Plasma Physics, D-85748 Garching, Germany\\
  ${}^3${}Max-Planck/Princeton-Center for Plasma Physics}

\ead{horvath.laszlo@reak.bme.hu}

\begin{abstract}
The present paper describes the effect of magnetic pick-up coil transfer functions on mode number analysis in magnetically confined fusion plasmas.
Magnetic probes mounted inside the vacuum chamber are widely used to characterize the mode structure of magnetohydrodynamic modes, as, due to their relative simplicity and compact nature, several coils can be distributed over the vessel.
Phase differences between the transfer functions of different magnetic pick-up coils lead to systematic errors in time- and frequency resolved mode number analysis.
This paper presents the first in-situ, end-to-end calibration of a magnetic pick-up coil system which was carried out by using an in-vessel driving coil on ASDEX Upgrade.
The effect of the phase differences in the pick-up coil transfer functions is most significant in the 50-250~kHz frequency range, where the relative phase shift between the different probes can be up to 1~radian $(\sim60^\circ)$.
By applying a correction based on the transfer functions we found smaller residuals of mode number fitting in the considered discharges.
In most cases an order of magnitude improvement was observed in the residuals of the mode number fits, which could open the way to investigate weaker electromagnetic oscillations with even high mode numbers.

% \vspace{1 cm}
% \noindent Revision number: \textbf{r\svnrev} - \svndate

\end{abstract}


\vspace{2pc}
\noindent{\it Keywords}: tokamak, Mirnov coil, magnetics, mode number, time-frequency analysis, Fourier transform, frequency transfer function% \hfill \red{Revision number: \textbf{r\svnrev}}
%\maketitle
% \ioptwocol

\section{Introduction}

Magnetohydrodynamic (MHD) instabilities have a strong influence on the plasma performance, therefore their thorough understanding is essential for stable tokamak operation.
Early steps of identification and analysis include determining the time-frequency evolution, spatial structure (characterized by mode numbers) and whether the perturbations are global or localized.
Several basic measurements of MHD modes rely on magnetic pick-up coils mounted on the inner side of the vacuum chamber.
Depending on the alignment these probes pick up different components of the magnetic field fluctuation.
On ASDEX Upgrade (AUG), Mirnov coils are placed on the vacuum vessel to measure the poloidal component of the magnetic field fluctuations $\tilde{B}_\mathrm{pol}$ and so-called ballooning coils are placed closer to the plasma on the low field side to measure the radial component $\tilde{B}_\mathrm{r}$.
Due to their relative simplicity and compact size, several coils can be distributed over different locations within the plasma vessel, making them invaluable candidates for the analysis of the mode structure.

Several methods are available to determine mode numbers of global MHD modes.
One example is the application of Singular Value Decomposition (SVD)~\cite{nardone92multichannel, kim99mhd}, which decomposes fluctuation data by choosing the basis vectors in the directions of maximum coherence.
With spatially distributed measurements one can obtain both the coherent spatial structure and the time evolution of a mode.
Another promising way of interpreting spatial and temporal data is the use of the Lomb periodogram, which was proposed by Zegenhagen and co-authors~\cite{zegenhagen06analysis}.
Our preferred approach, which is applied throughout this paper, is to apply linear continuous time-frequency transformations on the data and calculate the mode structure based on the phase differences between probe pairs~\cite{pokol08experimental, pokol10wavelet}.
A careful treatment of mode structure analysis takes into account the screening currents induced in the vessel wall and other in-vessel conducting components by a simulation of the perturbed magnetic field from an assumed current distribution on resonant surfaces~\cite{schittenhelm97analysis, igochine03investigation}.

Differences in the phase behaviour of the transfer functions corresponding to different coils can introduce systematic errors in the mode number fitting algorithms.
In order to correctly characterize mode numbers, each of the mentioned methods need to be evaluated on calibrated signals of the magnetic probes, which takes into account the effect of the magnetic coil response. 
Systematic errors come from various sources such as eddy currents, uncertainties in the probe positioning and alignment, resonances in the probe or the measurement circuit, due to the anti-aliasing analogue filter or even the analogue-digital converter (ADC).
We discovered systematic errors in the mode number analysis of strong MHD modes in the $50-250$~kHz frequency range on ASDEX Upgrade.
Several methods have been previously applied to model or measure the transfer function of the magnetic coils, however no in-situ, end-to-end calibration of the full diagnostic system using an in-vessel driving coil has been done before.
A remote calibration method has been developed on JET~\cite{heeter00fast, appel05calibration}.
This method determines the transfer function for the amplifier \& digitizer section by directly injecting a known signal.
The probe \& cable section of the system was modeled as a lumped-circuit and its transfer function was extracted from the frequency-dependent impedance measurement performed without entering the torus.
On TCV the response of the magnetic probes and the amplifying chain were measured separately before installation and a transfer function with two cut-off frequencies was fitted on the product of the two evaluated transfer functions~\cite{moret98magnetic, reimerdes01mhd}.
A Helmholtz coil was used to measure the transfer function of the coils outside the vacuum chamber on H-1NF heliac~\cite{haskey13multichannel}.
On TFTR, only the electronics of the Minov coils were calibrated~\cite{fredrickson95resutls}.

This paper presents the evaluation of the first in-situ, end-to-end calibration of a magnetic pick-up coil system which was carried out using an in-vessel driving coil~\cite{schmid05charakterisierung}.
A phase correction function was derived from the coil response measurements and it was applied to improve mode number determination on AUG.
We found that the effect of the phase differences in the pick-up coil transfer functions is most significant in the 50-250~kHz frequency range, where the relative phase shift between the different probes can be up to 1~radian $(\sim60^\circ)$.
The implications of this systematic error and the corresponding correction is demonstrated on strong Toroidicity-induced Alfv\'{e}n Eigenmodes (TAEs) observed in AUG discharges, where a significant improvement in toroidal mode number fitting is achieved.

The paper is organized as follows.
In section~\ref{sec:tools} we present the mathematical background of our mode number analysis and the set of magnetic pick-up coils used.
Systematic errors observed in magnetic measurements are described in section~\ref{sec:systematic}.
Section~\ref{sec:resp-meas} details the in-situ transfer function measurements on ASDEX Upgrade.
The implementation of the correction, including examples, is described in section~\ref{sec:correction}; followed by the conclusions in section~\ref{sec:conc}.

\section{Analysis principles and measurement set-up}\label{sec:tools}

Our mode number analysis uses linear continuous time-frequency transforms~\cite{mallat08wavelet}. These are calculated by expanding the signal $f(t)$ on the basis of families of so-called time-frequency atoms:
\begin{equation}\label{eq:transform}
  T f(t,\omega) = \langle f, g_{t, \omega} \rangle = \int\limits_{-\infty}^{+\infty} f(t') g^*_{t, \omega}(t') \dt'\ ,
\end{equation}
where $g_{t, \omega}$ is a time-frequency atom, whose energy is well localized in both time and frequency.
Variables $t$ and $\omega$ are the time and frequency indices of the atom identifying its position on the time-frequency plane, and the $^*$ represents the complex conjugation.
The energy density distribution can then be calculated by taking the absolute value squared~\cite{mallat08wavelet}:
\begin{equation}\label{eq:energy_density}
  E f(t, \omega) = |T f(t, \omega)|^2\ .
\end{equation}
Our mode number calculation process is based on the phase of the cross-transform:
\begin{equation}\label{eq:cross-phase}
  \varphi_{kl}^{\mathrm{meas}} (t, \omega) = \arg\left\{T f_k(t, \omega) T f_l^*(t, \omega)\right\} \ ,
\end{equation}
where $f_k$ and $f_l$ represent the signals of probes placed in different positions.

In this paper, we demonstrate the mode number determination method by using Short Time Fourier Transform (STFT)~\cite{mallat08wavelet}.
Note that this method also works with continuous analytical wavelet transform \cite{mallat08wavelet}.
STFT is a continuous time-frequency transform where the family of time-frequency atoms are generated by shifting a real and symmetric window $g(t')$ in time $(t)$ and frequency $(\omega)$:
\begin{equation}\label{eq:stft-atom}
  g_{t, \omega}(t') = e^{\rmi\omega t'} g(t' - t) \ ,
\end{equation}
which gives a constant time-frequency resolution on the time-frequency plane.
Following from \eqref{eq:transform} STFT can be written as
\begin{eqnarray}\label{eq:stft}
  S f(t, \omega) &= \langle f, g_{t, \omega} \rangle = \nonumber\\
  &= \int\limits_{-\infty}^{+\infty} f(t') e^{-\rmi \omega t'} g(t' - t) \dt' \ ,
\end{eqnarray}
where $g(t')$ is a Gaussian function in our case.
The energy density distribution defined in \eqref{eq:energy_density} calculated from STFT is called a \emph{spectrogram}, which is ideal to investigate the time-frequency evolution of non-stationary modes~\cite{sertoli13characterization, pokol13continuous}.
In order to apply a continuous transform on discrete time signals, the transform has to be discretized in a way to avoid the degradation of the time-shift invariance property of the transform~\cite{pokol07application}.

\begin{figure}[htb!]\centering
  \includegraphics[width = 70mm]{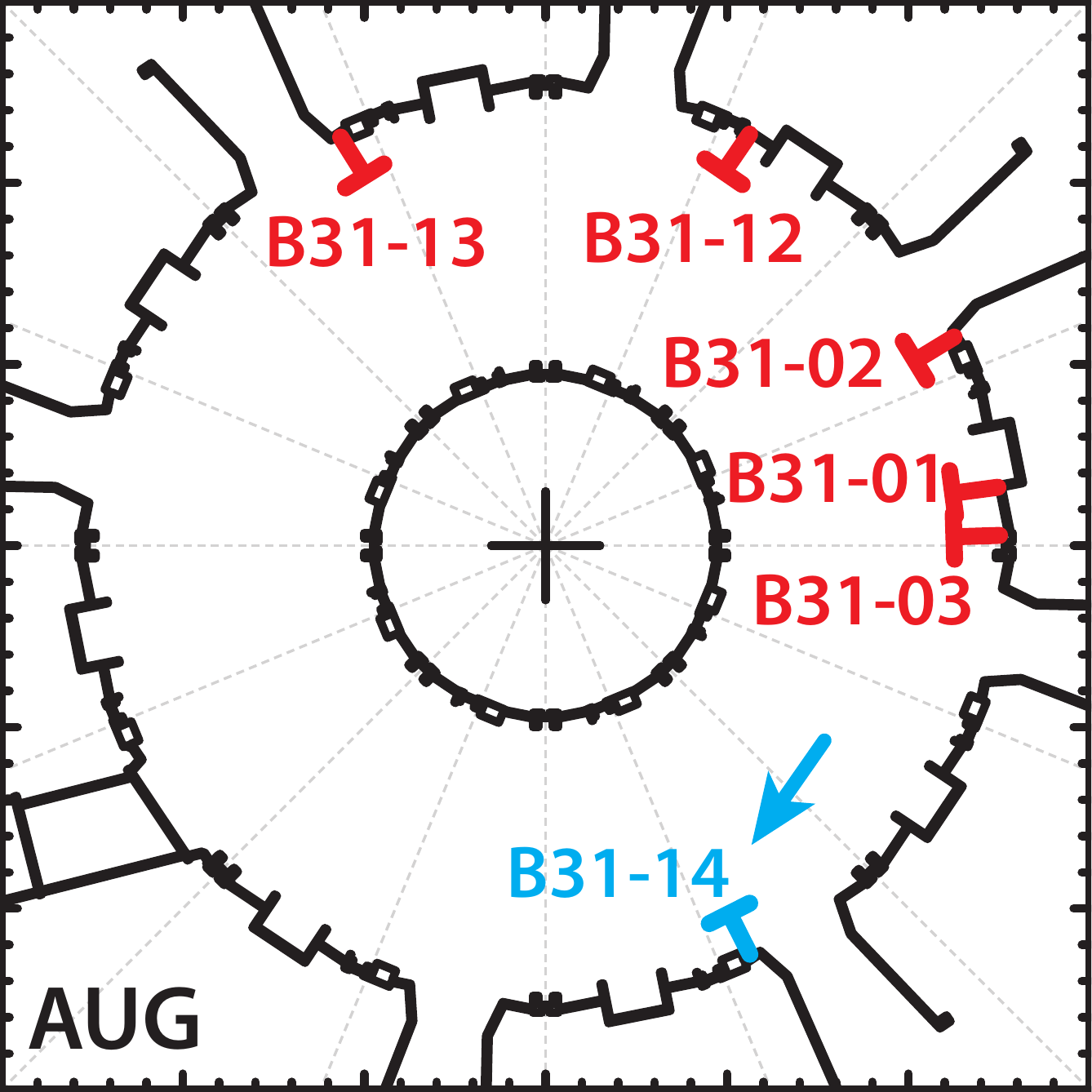}
  \caption{A top-down view of ASDEX Upgrade indicating the position of the ballooning coils in the toroidal array. The arrow marks the B31-14 probe which is used as a reference.}
  \label{fig:probe_position}
\end{figure}

Hereinafter, we focus on toroidal mode number analysis on AUG, derived from the measurements of six ballooning coils placed close to the separatrix ($\sim10$~cm) at the outer midplane in different toroidal, but almost identical poloidal positions.
These ballooning coils are rectangular spirals (etched out vacuum-compliant printed circuit boards) which measure the changes in the radial component of the magnetic field.
The vanishing dimension in the direction of the field makes them extremely sensitive for short wavelength perturbations.
The toroidal positioning of these probes is illustrated in figure~\ref{fig:probe_position}.
The measurement set-up on AUG contains ADCs which have a $2$ MHz sampling frequency and they include a 3dB low pass Bessel filter with $512$~kHz cut-off~\cite{wenninger13nonlinear}.

The measurement of magnetic field fluctuations produced by MHD modes allows the reconstruction of their mode structure.
Harmonics of global MHD eigenmodes are generally assumed in the following form~\cite{zegenhagen06analysis}:
\begin{eqnarray}\label{eq:eigenmode}
  & A_{m,n}(\rho,\theta^*,\phi,t) = \nonumber\\
  & = A_{m,n}(\rho,\theta^*)\exp\{\rmi(m\theta^*+n\phi-\omega t)\} \ ,
\end{eqnarray}
where $A$ is the observable quantity (in our case the radial magnetic field perturbation $\tilde{B}_\mathrm{r}$) as a function of time $t$ and the so-called straight field line coordinates $(\rho,\theta^*,\phi)$~\cite{schittenhelm97analysis}, $A(\rho,\theta^*)$ is the radial eigenfuntion and $\omega$ is the mode frequency.
The mode structure of the harmonic is characterized by the $m$ poloidal and $n$ toroidal mode numbers.
As follows from \eqref{eq:eigenmode}, in the case of a single, pure sinusoidal, globally coherent mode at fixed $(\rho,\theta^*,t)$, but at different toroidal angles, the relative phase between $A_{m,n}(\phi_k)$ and $A_{m,n}(\phi_l)$ is proportional to the relative toroidal angle $\phi_k - \phi_l$ and the ratio is the toroidal mode number $n$:
\begin{eqnarray}\label{eq:proportional}
  \varphi_{kl}^{\mathrm{id}}{}_{\big|_{\theta_k = \theta_l}} &= 
  \arg\left\{ A_{m,n}(\phi_k) \right\} - \arg\left\{ A_{m,n}(\phi_l) \right\} = \nonumber\\
  &= n(\phi_k - \phi_l) \ .
\end{eqnarray}
Thus, in order to characterize the spatial structure of the mode, one has to determine its phase in different locations $(\phi_i)$.
Equation~\eqref{eq:proportional} shows that the slope of the linear curve, fitted on relative phases between all pairs of signals as a function of the relative probe position, gives the toroidal mode number $n$.
Our method searches for the integer mode number value where the residual $Q$ of the fit is minimal:
\begin{equation}\label{eq:residual}
  Q = \sum_{\mathcal{P}} \| \varphi_{\mathcal{P}} - n\phi_{\mathcal{P}} \|^2 \ ,
\end{equation}
where $\varphi_\mathcal{P}$ is the phase between pairs of signals, $\phi_\mathcal{P}$ is the relative probe position, $n$ is the toroidal mode number, the sum is executed on the $\mathcal{P}$ signal pairs and $\| . \|$ is the norm obtained by taking the optimum shift of $\varphi_{\mathcal{P}}$ by $2\pi z$, where $z$ is an integer number.
The sign of the mode number determines the direction of propagation in the device frame.
If the helicity (which is determined by the direction of the toroidal magnetic field and the plasma current) is known, the sign of the mode number can be related to the ion or electron diamagnetic drift direction.

A time-frequency resolved map of the mode numbers can be generated by performing the linear fit in each time-frequency point~\cite{pokol08experimental}.
Note that this result is only valid, where global modes exist.
The time-frequency based mode number analysis presented above is implemented in NTI Wavelet Tools custom data processing toolbox\footnote{Developed and maintained in the Institute of Nuclear Techniques (NTI), Budapest University of Technology and Economics, \url{https://deep.reak.bme.hu/projects/wavelet}}.

In principle, the calculated cross-phase in \eqref{eq:cross-phase} is a reasonable approximation of the relative phase of the mode, however, if the phase information carried by the measured signals suffer from systematic errors, the mode number analysis can yield erroneous results.

\section{Systematic errors in magnetic measurements}\label{sec:systematic}

\begin{figure}[htb!]\centering
  \includegraphics[width = 80mm]{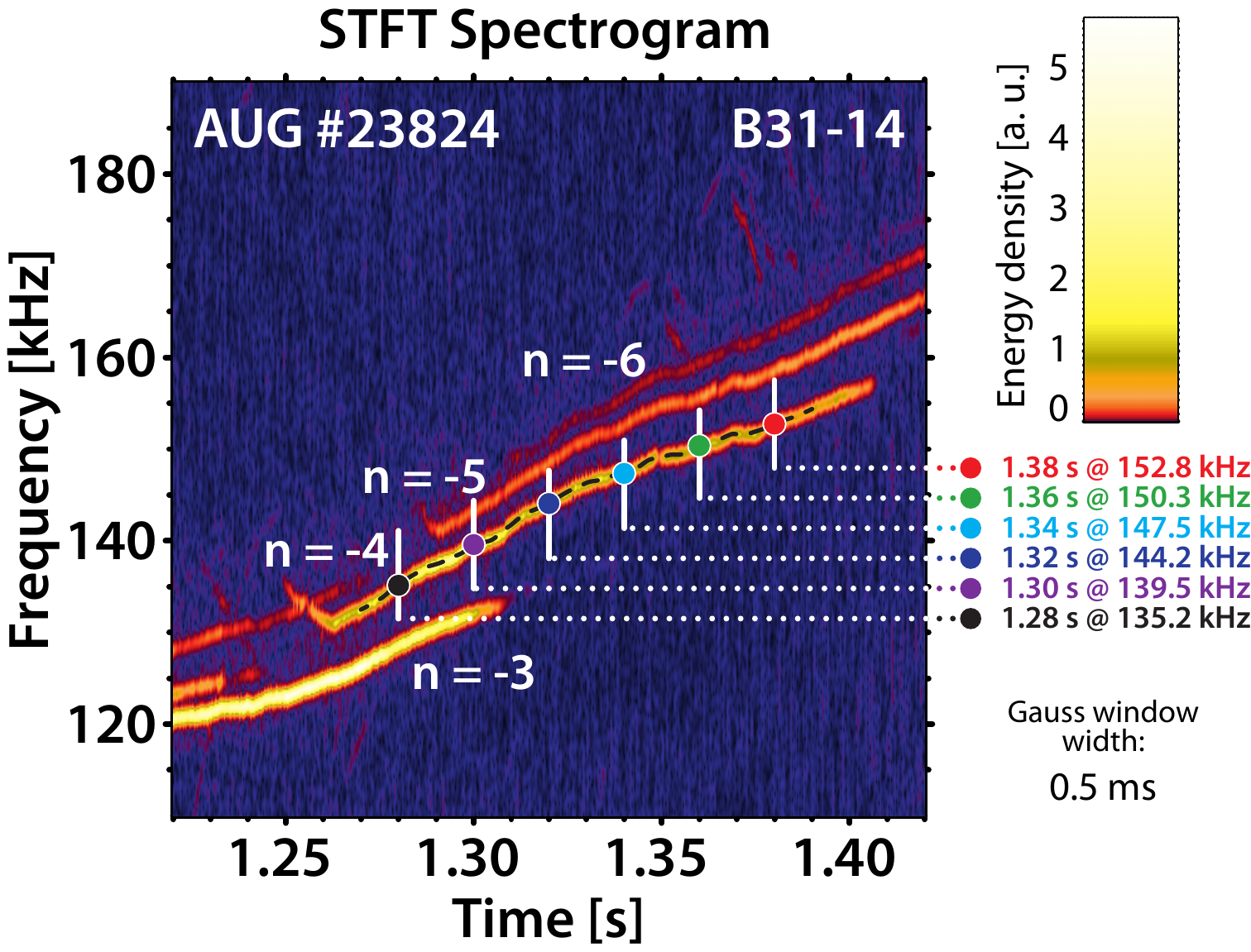}
  \caption{Magnetic spectrogram of strong coherent TAE harmonics with different toroidal mode numbers. The frequency of the mode with $n = -4$ toroidal mode number (as calculated in figure~\ref{fig:phasediagram_no}) is indicated with black dashed line. The time-frequency points indicated with colours were chosen for mode number analysis.}
  \label{fig:spectrogram}
\end{figure}

The systematic errors can be easily recognized in the presence of strong, coherent MHD modes.
In such a case, one would expect that the calculated relative phases (eq.~\eqref{eq:cross-phase}) between all pairs of signals would lie on a straight line as a function of the relative probe position.
As we demonstrate in the following example, systematic deviations from the best fitting line were observed.
In our example we follow the time evolution of a strong TAE illustrated in figure~\ref{fig:spectrogram}.
Since 6 probes are available in the toroidal array, 15 pairs of coils can be selected.

\begin{figure}[htb!]\centering
  \includegraphics[width = 150mm]{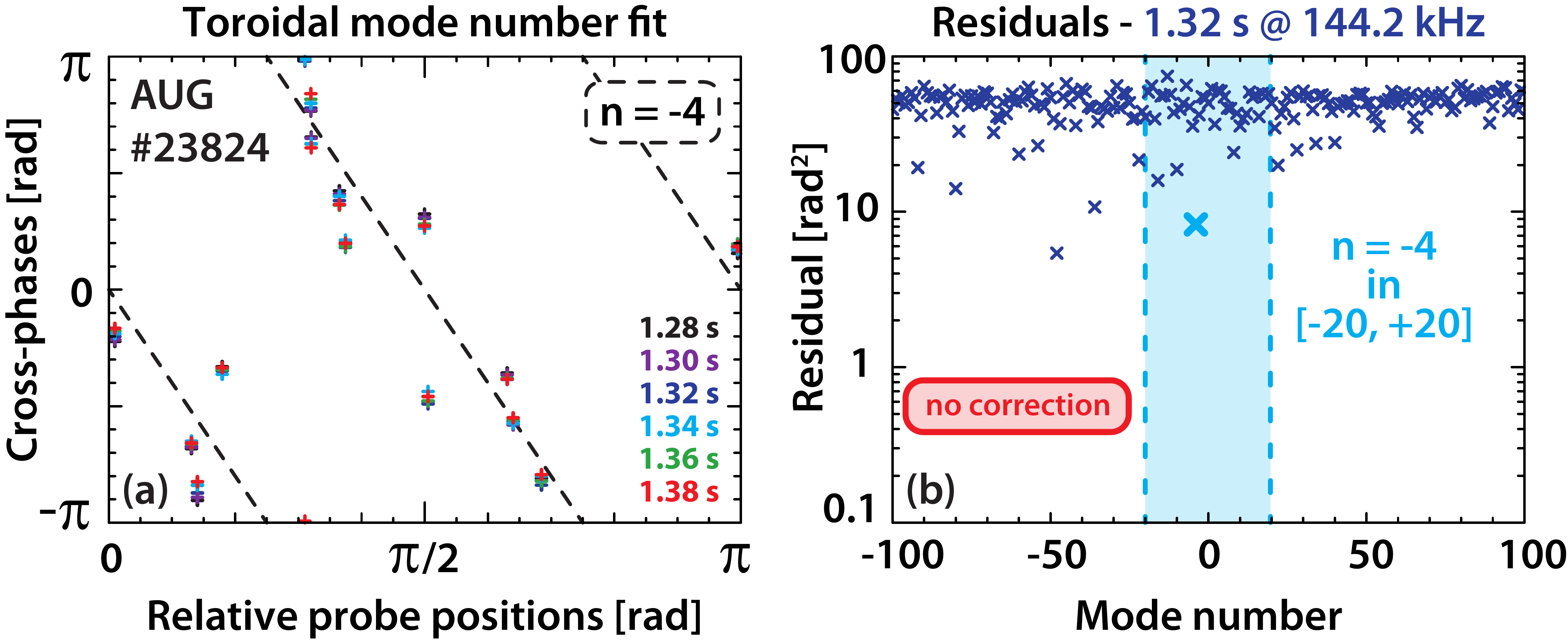}
  \caption{\textbf{(a)} The relative phases between all pairs of signals are plotted as a function of relative probe position.  The different colours present the different time points depicted in figure~\ref{fig:spectrogram}. The best fitting line with $n = -4$ slope is plotted with black solid lines. For the helicity of this discharge, the negative mode number means that the mode propagates in the ion diamagnetic drift direction. \textbf{(b)} The residuals (defined in \eqref{eq:residual}), which quantify the differences between the fit and the measurements, are shown as a function of the mode number. In the $[-20, +20]$ range $n = -4$ is the best fitting value.}
  \label{fig:phasediagram_no}
\end{figure}

\begin{figure}[htb!]\centering
  \includegraphics[width = 80mm]{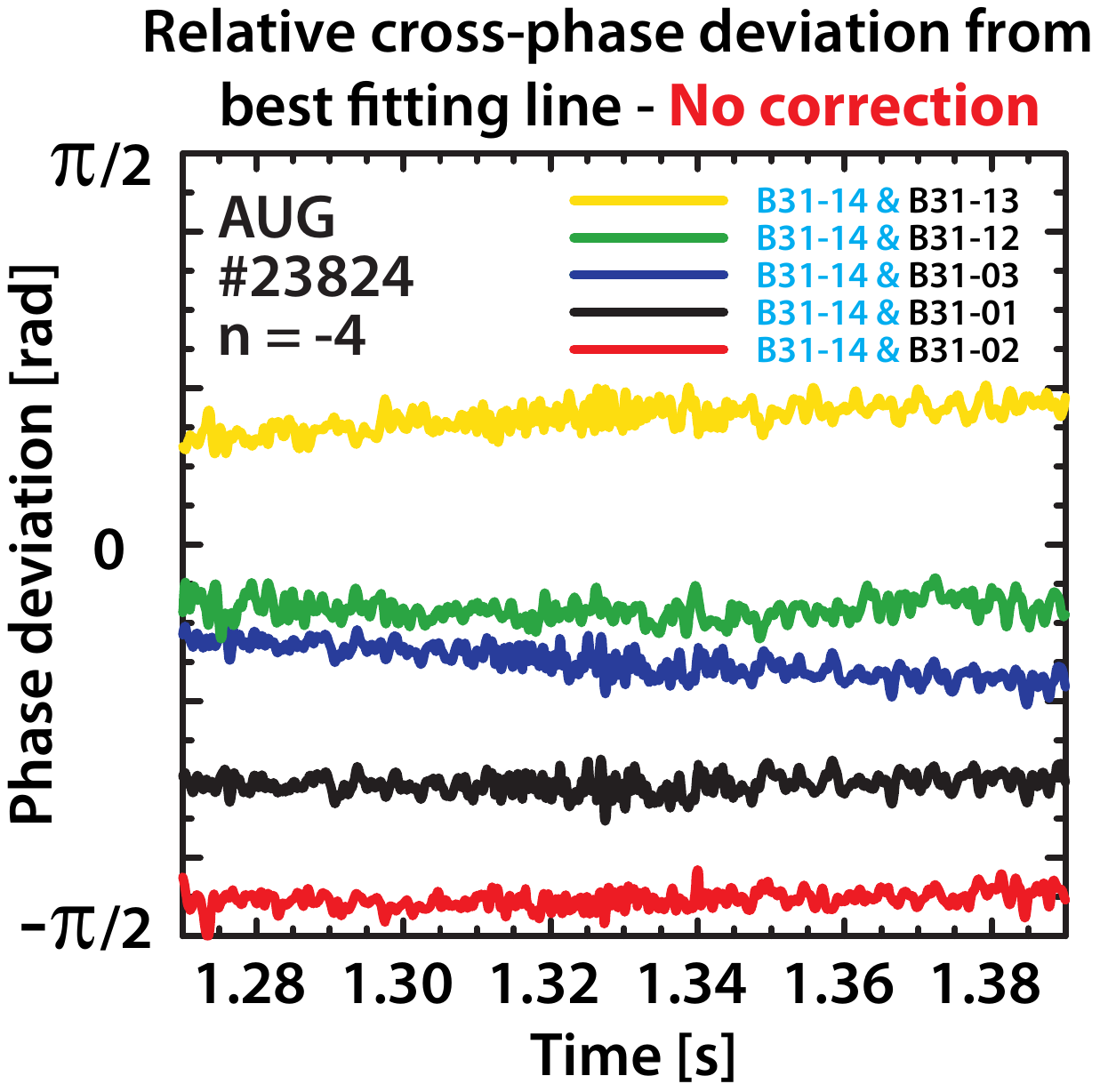}
  \caption{The deviation of the relative phases from the $n = -4$ for independent channel pairs as a function of time. The systematic behaviour of the deviation is clearly visible.}
  \label{fig:deviation_no}
\end{figure}

We trace the time evolution of the mode frequency using a ridge-following algorithm~\cite{papp11low, horvath14changes}, the result of which is called the ridge of the STFT transform.
The frequency ridge of the mode with toroidal mode number $n = -4$ is indicated with dashed black line on figure~\ref{fig:spectrogram}.
We mark the time-frequency points selected for mode number analysis by the cross section of the time points indicated with white vertical lines on figure~\ref{fig:spectrogram} and the ridge of the mode.
The relative phases calculated between all signal pairs are plotted as a function of relative probe position on figure~\ref{fig:phasediagram_no}a.
The different colours represent the time points corresponding to figure~\ref{fig:spectrogram}.
The linear fit on the relative phases is performed in each time instant trying to fit integer mode numbers from the interval $[-100, +100]$.
The residuals (defined in \eqref{eq:residual}) of the fit at $1.32$ sec are shown as a function of the mode number in figure~\ref{fig:phasediagram_no}b.
In the interval $[-20, +20]$ the best fitting toroidal mode number is $n=-4$, which also holds in the other investigated time points.
However, we observe a systematic deviation of the relative phases from the $n = -4$ line (plotted with black solid lines on figure~\ref{fig:phasediagram_no}a).
The deviation of the relative phases from the $n = -4$ for independent channel pairs as a function of time is shown in figure~\ref{fig:deviation_no}.
Such systematic deviations were observed in all investigated cases.

Such systematic errors are likely to arise from the differences in the phase behaviour of the transfer functions corresponding to different magnetic coils.
In the next section we present the results and implications of in-situ magnetic coil response measurements performed on ASDEX Upgrade.

\section{Coil response measurement on ASDEX Upgrade}\label{sec:resp-meas}

We modeled the measurement circuit of a magnetic probe as a linear single-input single-output (SISO) system.
The input of the system is in principle the time derivative of the radial magnetic field in the centre of a ballooning coil and the output is the signal registered by the ADC.
The frequency-dependent transfer function of a linear system at a given frequency $\omega$ is
\begin{equation}\label{eq:transfer}
  H(\omega) = \frac{O(\omega)}{I(\omega)} \ ,
\end{equation}
where $O(\omega)$ and $I(\omega)$ are the Fourier transforms of the output and input signals.

In order to determine the transfer function of the ballooning coils an in-situ measurement was performed on ASDEX Upgrade~\cite{schmid05charakterisierung}.
Since this measurement was carried out inside the fully equipped tokamak device on the installed magnetic probes, the result contains the effect of the coil response, the analogue filter (a low pass, 3dB Bessel filter with $512$~kHz cut-off), the ADC ($2$~MHz sampling frequency), the signal transmission cables and to some extent even the effect of eddy currents in the in-vessel structures.

\begin{figure}[htb!]\centering
  \includegraphics[width = 70mm]{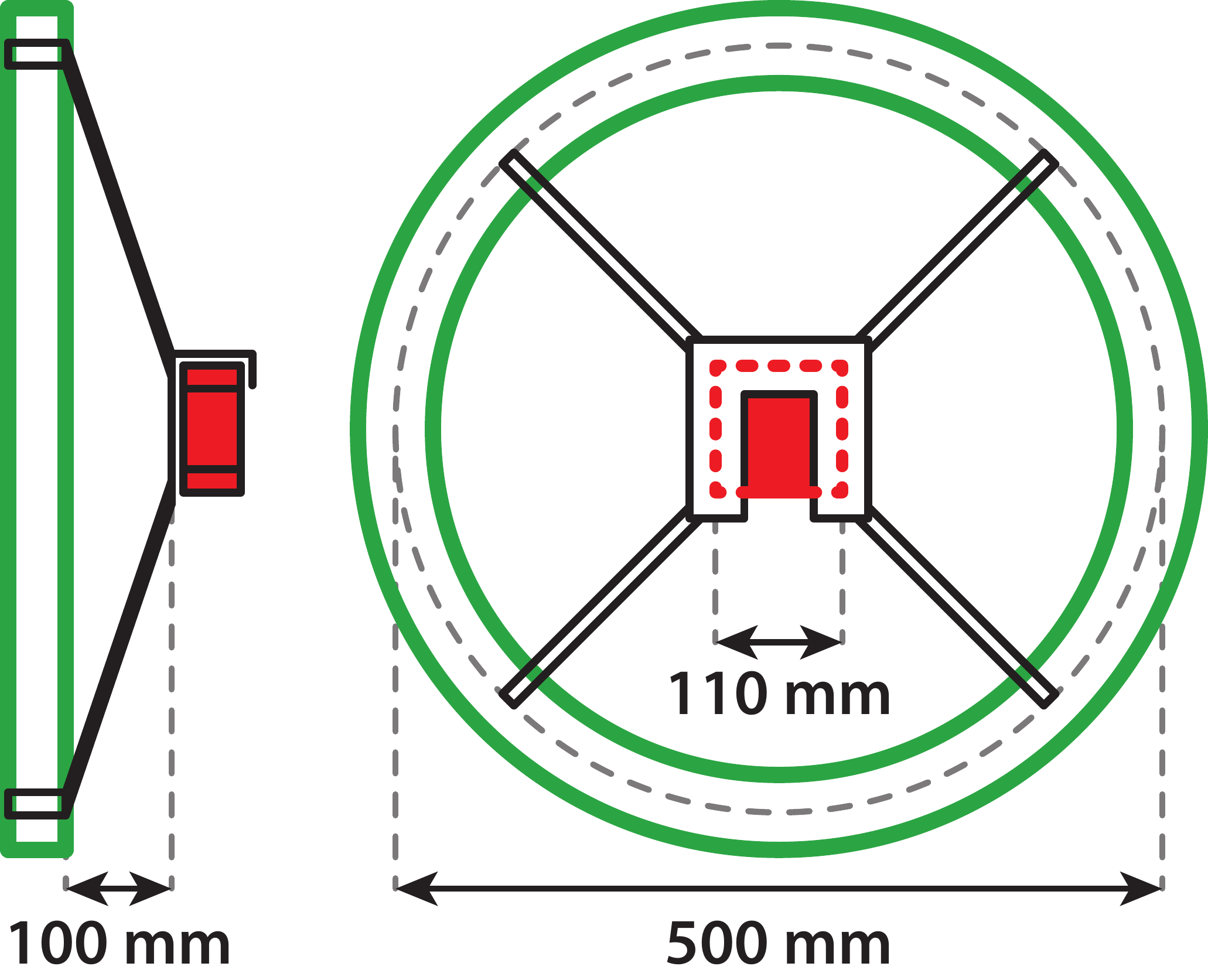}
  \caption{The geometry of the measurement set-up from which the transfer functions of the magnetic coils are estimated. The red pick-up coil was excited by the green driving coil. The black supporting structure provided the consistency of the measurement geometry from probe to probe.}
  \label{fig:driving_coil}
\end{figure}

\begin{figure}[htb!]\centering
  \includegraphics[width = 120mm]{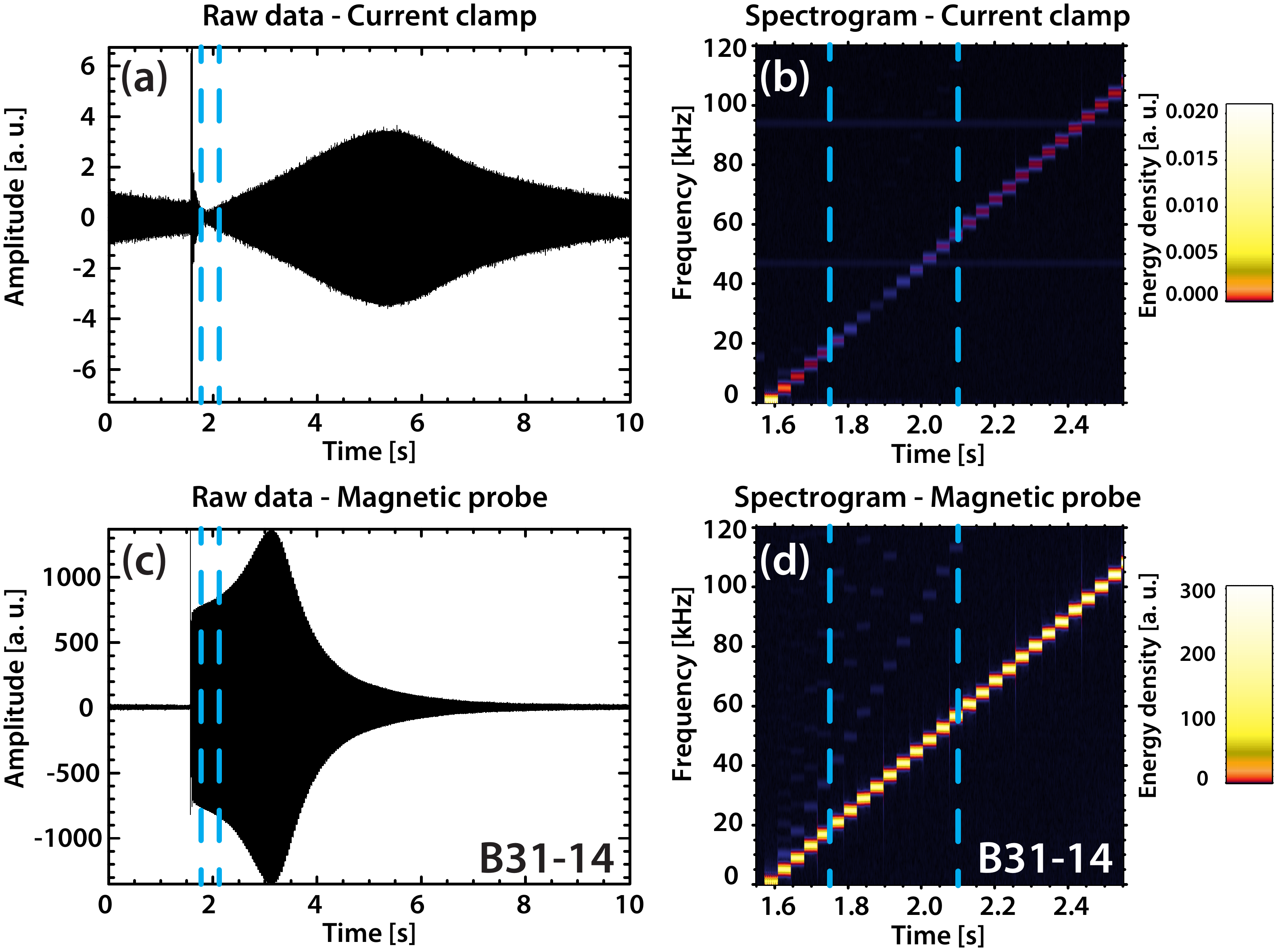}
  \caption{Raw signals and spectrograms of the signals of transfer function measurements. \textbf{(a)} The raw signal of the current clamp measurement of the B31-14 coil. \textbf{(b)} The spectrogram of a part of the current clamp measurement. \textbf{(c)} The raw signal of the B31-14 coil. \textbf{(d)} The spectrogram of a part of the magnetic probe signal.}
  \label{fig:resonance}
\end{figure}

The measurements from which the transfer functions of the magnetic coils are estimated were carried out by exciting the magnetic probes one by one with a driving coil placed in front of the probes.
The geometry of the measurement set-up is shown in figure \ref{fig:driving_coil}.
The input voltage of the driving coil came from a constant amplitude function generator, that swept the $1$~kHz~--~$1$~MHz frequency range in $4$~kHz steps with a 9 second repetition time.
The current flowing in the driving coil was measured with a current clamp (clamp-on AC current meter).
The magnetic field at the probe is directly related to the current ($\mathcal{I}_\mathrm{d}$) flowing in the driving coil:
\begin{equation}\label{eq:magnetic}
  B = c_{\mathrm{d}} \mathcal{I}_\mathrm{d} \ ,
\end{equation}
where $c_{\mathrm{d}}$ is a constant which gives the strength of the magnetic field per unit current and has a dimension $T/A$.
Its value is unequivocally determined by the geometric structure of the measurement set-up.
Since the magnetic probe detects the changes of the magnetic field, we need the time derivative of \eqref{eq:magnetic}, which corresponds to a multiplication by $\rmi\omega$ in Fourier space (with $\rmi$ being the imaginary unit).
The signals of the magnetic probes were recorded using the shotfile system, exactly the same way as during a plasma discharge.
The measurement of the input and output signals was started at random phases during the sweep, but the 10 second recording length ensured that the whole frequency range is covered.
The raw signal of one of the current clamp measurements is shown in figure~\ref{fig:resonance}(a).
Figure~\ref{fig:resonance}(b) shows the spectrogram for a part of the signal.
In this case the sweep started at about $t = 1.6$~sec and the $4$~kHz frequency step is clearly visible on the spectrogram.
Similarly, the raw signal of the magnetic probe and the corresponding spectrogram is shown in figures~\ref{fig:resonance}(c-d).

The length of a constant frequency period is about $36$~ms.
For the purpose of the present evaluation a $20$~ms long interval was selected from the middle of the constant frequency periods to safely exclude the frequency jumps.
Each $20$~ms long interval was handled as a separate measurement at a given frequency.
The exact frequency of the driving signal during the constant frequency periods was measured by finding the spectral maximum of each $20$~ms long signals.

If the changes of the magnetic field at the coil would be known, the ideal transfer function could be calculated in the following way:
\begin{equation}\label{eq:ideal_transfer}
  H^{\mathrm{id}}(\omega) = \frac{O(\omega)}{\rmi\omega c_\mathrm{d} B(\omega)} \ ,
\end{equation}
where $B(\omega)$ denotes the Fourier transform of the magnetic field at the coil, $O(\omega)$ denotes the Fourier transform of the output signal registered by the ADC of the magnetic coil.
However, $B(\omega)$ is not know, only the signal of the current clamp is registered which is denoted by $I(\omega)$ since this signal is considered as the input of the measurement.
The relation between the magnetic field at the coil ($B(\omega)$) and the signal of the current clamp ($I(\omega)$) is given by the transfer function of the current clamp measurement system:
\begin{equation}\label{eq:clamp_transfer}
  H^{\mathrm{cl}}(\omega) = \frac{I(\omega)}{B(\omega)} \ .
\end{equation}
Since $H^{\mathrm{cl}}(\omega)$ is not known,
\begin{equation}\label{eq:meas_transfer}
  H^{\mathrm{meas}}(\omega_j) = \frac{O(\omega_j)}{\rmi\omega I(\omega_j)} \ ,
\end{equation}
function was calculated separately for the constant $\omega_j$ frequency periods which, when assembled, gave a first approximation for transfer function of a given magnetic coil at the discrete $\omega_j$ frequencies.

\begin{figure}[htb!]\centering
  \includegraphics[width = 150mm]{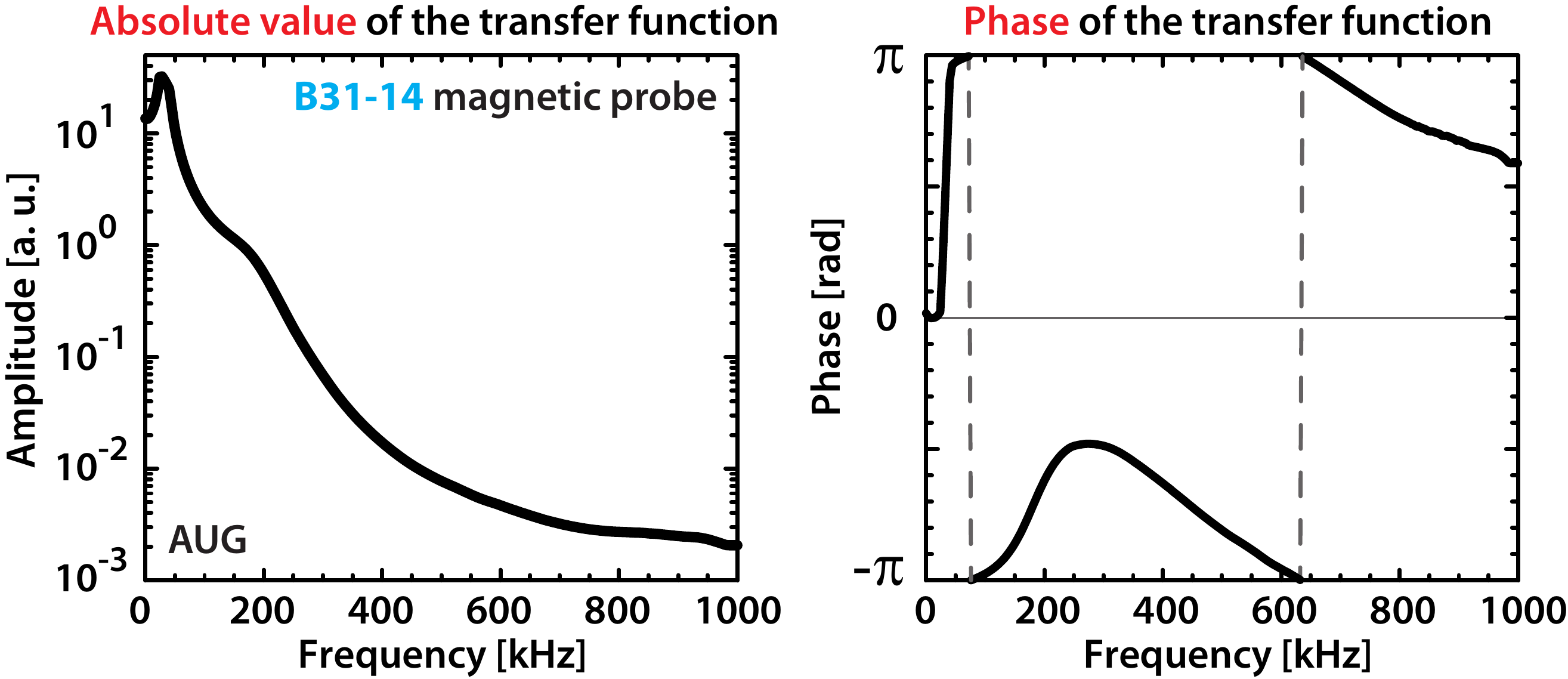}
  \caption{The absolute value and phase of the calculated transfer function of the B31-14 coil.}
  \label{fig:abstransfer}
\end{figure}

The absolute value and phase of $H^{\mathrm{meas}}(\omega_j)$ (defined in \eqref{eq:meas_transfer}) of coil B31-14 is shown in figure~\ref{fig:abstransfer}.
The absolute value changes more than four orders of magnitudes over the $0-1$~MHz frequency range.
It has a peak between $20$~kHz and $40$~kHz which arises from the fact that despite of the high amplitude on the magnetic coil, the signal of the current clamp strongly decreased in this frequency range.
This phenomenon is visible both on the raw signals and on the spectrograms in figure~\ref{fig:resonance} where the relevant time interval is indicated with dashed blue lines.
Such behaviour of the magnetic coils has not been recognized in plasma discharges which suggests that the driving magnetic field strength was not measured accurately in this frequency range.
This low frequency resonance is not the feature of the magnetic coil, but it is most probably due to a resonance in the current clamp measurement set-up.
Since $H^{\mathrm{cl}}(\omega)$ (defined in \eqref{eq:clamp_transfer}) is not known, but it has a significant impact on the results, the absolute calibration of the coils is not possible from the recorded data.
However, the response of the different coils was measured by using the same setting, thus the measurement is excellent for relative calibration which is shown in the next part of this section.

In order to prove the pertinence of the relative calibration, the relation between the ideal transfer function of the system (defined in \eqref{eq:ideal_transfer}) and the transfer function defined in \eqref{eq:meas_transfer} has to be examined.
Using \eqref{eq:ideal_transfer} and \eqref{eq:clamp_transfer} we define the relative transfer function between two different magnetic coils in the following way:
\begin{eqnarray}\label{eq:ideal_transfer1}
  H_{kl}(\omega) &= \frac{H^{\mathrm{id}}_k(\omega)}{H^{\mathrm{id}}_l(\omega)} =
  \frac{O_k(\omega) / \rmi\omega c_{\mathrm{d}} B_k(\omega)}{O_l(\omega) / \rmi\omega c_{\mathrm{d}} B_l(\omega)} = \nonumber\\
  &= \frac{ O_k(\omega) / I_k(\omega) / H_k^{\mathrm{cl}}(\omega) }
  { O_l(\omega) / I_l(\omega) / H_l^{\mathrm{cl}}(\omega) } \ .
\end{eqnarray}
Since the same set-up was used for each measurement, the transfer function of the current clamp is the same everywhere:
$H_k^{\mathrm{cl}}(\omega) = H_l^{\mathrm{cl}}(\omega) = H^{\mathrm{cl}}(\omega)$.
Thus the relative transfer function can be determined without the knowledge of $H^{\mathrm{cl}}(\omega)$:
\begin{equation}\label{eq:ideal_transfer2}
  H_{kl}(\omega) = \frac{ O_k(\omega) / I_k(\omega) }{ O_l(\omega) / I_l(\omega) } \ .
\end{equation}

\begin{figure}[htb!]\centering
  \includegraphics[width = 80mm]{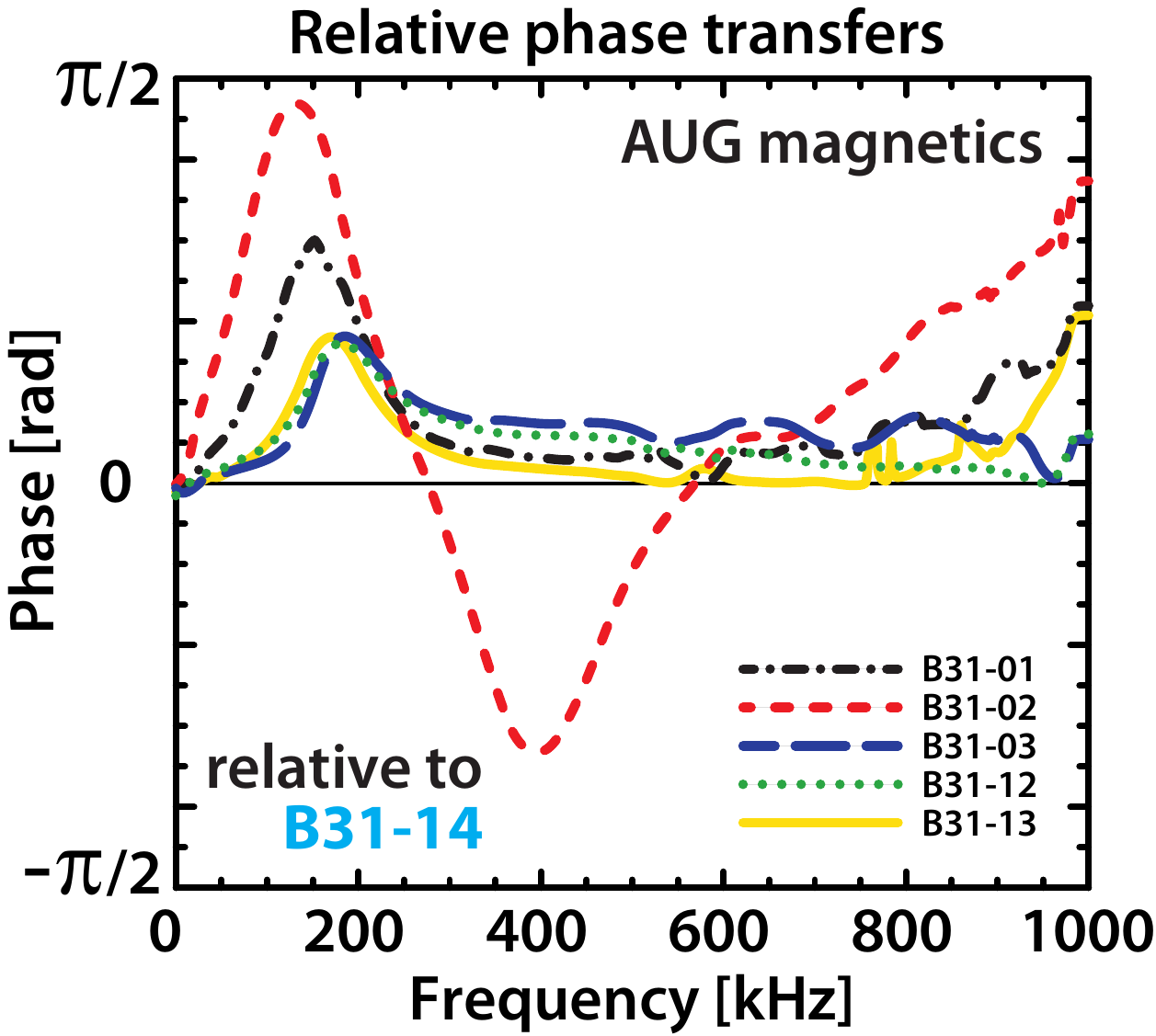}
  \caption{The relative phase difference between the transfers functions of ballooning coils relative to  B31-14.}
  \label{fig:relative_phase}
\end{figure}

The measurement has been performed on each ballooning coil and their relative transfer functions defined in \eqref{eq:ideal_transfer2} between all ballooning coil pairs were calculated.
The phases of the relative transfer functions of independent pairs relative to B31-14 are shown in figure~\ref{fig:relative_phase}.
The effect caused by the differences in the individual pick-up coil transfer functions is most significant in the 50-250~kHz frequency interval, which is the range of ICRH ion-driven TAE modes and other fast particle driven Alfv\'{e}nic instabilities on AUG~\cite{garciamunoz11fast}.
In this interval, the relative phase delay between the different probes can reach 1~radian $(\sim60^\circ)$.
The significant variation in the transfer functions between the various probes is most probably caused by effects of the induced mirror currents within the in-vessel structures, which is not easily accessible theoretically and by the different resistance of the probes (from $57.9$ $\Omega$ to $191.4$ $\Omega$).

\section{Correction for the coil response in mode number analysis}\label{sec:correction}

The relative phase differences presented in section~\ref{sec:systematic} can be corrected in the mode number analysis by using the relative transfer function defined in \eqref{eq:ideal_transfer2}.
This correction significantly reduces the systematic errors recognized in the phase of the magnetic probe signals.

For a pure sinusoidal mode structure, the relative phase of the mode described by \eqref{eq:proportional} is equivalent to the short time Fourier cross-transform (defined in \eqref{eq:cross-phase}) calculated from the magnetic field fluctuations:
\begin{equation}\label{eq:corr1}
  \varphi_{kl} = \arg\left\{S b_k(t, \omega) S b_l^*(t, \omega)\right\} \ ,
\end{equation}
where $S b(t, \omega)$ is the short time Fourier transform of the magnetic field in the centre of one coil.
If $S b_l(t, \omega)$ is non-zero - which is a necessary condition - then:
\begin{eqnarray}\label{eq:corr2}
  \varphi_{kl} &= \arg\left\{S b_k(t, \omega) \frac{|S b_l(t, \omega)|^2}{S b_l(t, \omega)}\right\} = \nonumber\\
  &= \arg\left\{ \frac{S b_k(t, \omega)}{S b_l(t, \omega)}\right\} \ .
\end{eqnarray}
Recalling \eqref{eq:ideal_transfer} and considering \eqref{eq:ideal_transfer1} and \eqref{eq:ideal_transfer2} $\varphi_{kl}$ can be expressed with the $S o(t, \omega)$ short time Fourier transforms of the measured signals and the $H_{kl}(\omega)$ relative transfer function:
\begin{eqnarray}\label{eq:corr3}
  \varphi_{kl} &= \arg\left\{ \frac{S o_k(t, \omega) / \rmi\omega H_k^{\mathrm{id}}(\omega) }
  { S o_l(t, \omega) / \rmi\omega H_l^{\mathrm{id}}(\omega) }\right\} = \nonumber\\
  &= \arg\left\{ \frac{ S o_k(t, \omega)}{S o_l(t, \omega)} \frac{1}{H_{kl}(\omega)} \right\} \ ,
\end{eqnarray}
which means that applying the correction of the relative transfer functions of the magnetic probes evaluated in section \ref{sec:resp-meas},  the $\varphi_{kl}$ cross-phase of the magnetic fluctuations can be more accurately determined from the measured signals.

\begin{figure}[htb!]\centering
  \includegraphics[width = 150mm]{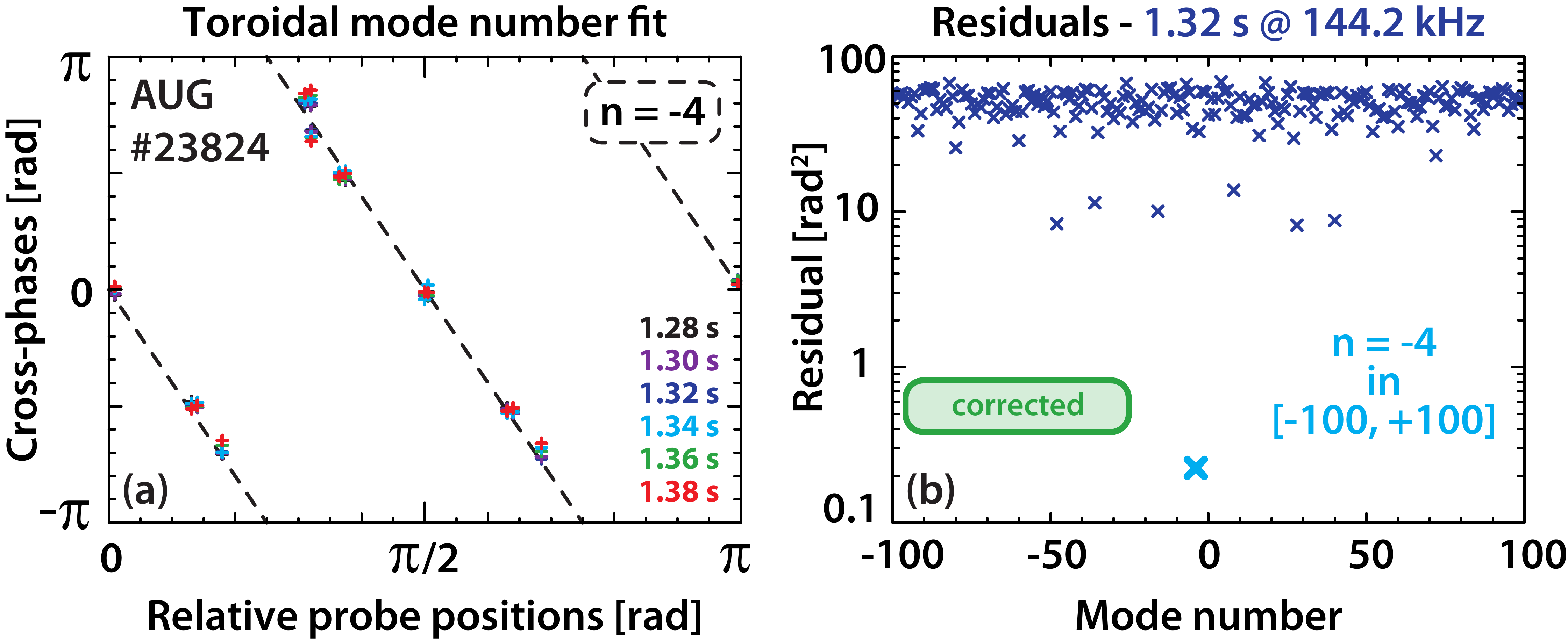}
  \caption{\textbf{(a)} The relative phases between all pairs of signals are plotted as a function of relative probe position. The different colours represent the different time points as shown on figure~\ref{fig:spectrogram}. The best fitting line with $n=-4$ slope is plotted with black solid lines. Compared to figure~\ref{fig:phasediagram_no}, here the cross-phases were corrected with the values derived from the coil response measurement. \textbf{(b)} The residuals (defined in \eqref{eq:residual}), which quantify the differences between the fit and the measurements, are shown as a function of the mode number. Compared to figure~\ref{fig:phasediagram_no}b the $n=-4$ mode number fits very much better.}
  \label{fig:phasediagram_corr}
\end{figure}

The toroidal mode number of the TAE presented in section~\ref{sec:systematic} is evaluated again using \eqref{eq:corr3}.
The resulting toroidal mode number of $n=-4$ agrees with the result achieved without the correction.
However, the correction leads to a significantly better fit, as illustrated in figure~\ref{fig:phasediagram_corr}a and quantitatively characterized by the differences between the residuals shown in figures~\ref{fig:phasediagram_no}b~and~\ref{fig:phasediagram_corr}b.

\begin{figure}[htb!]\centering
  \includegraphics[width = 80mm]{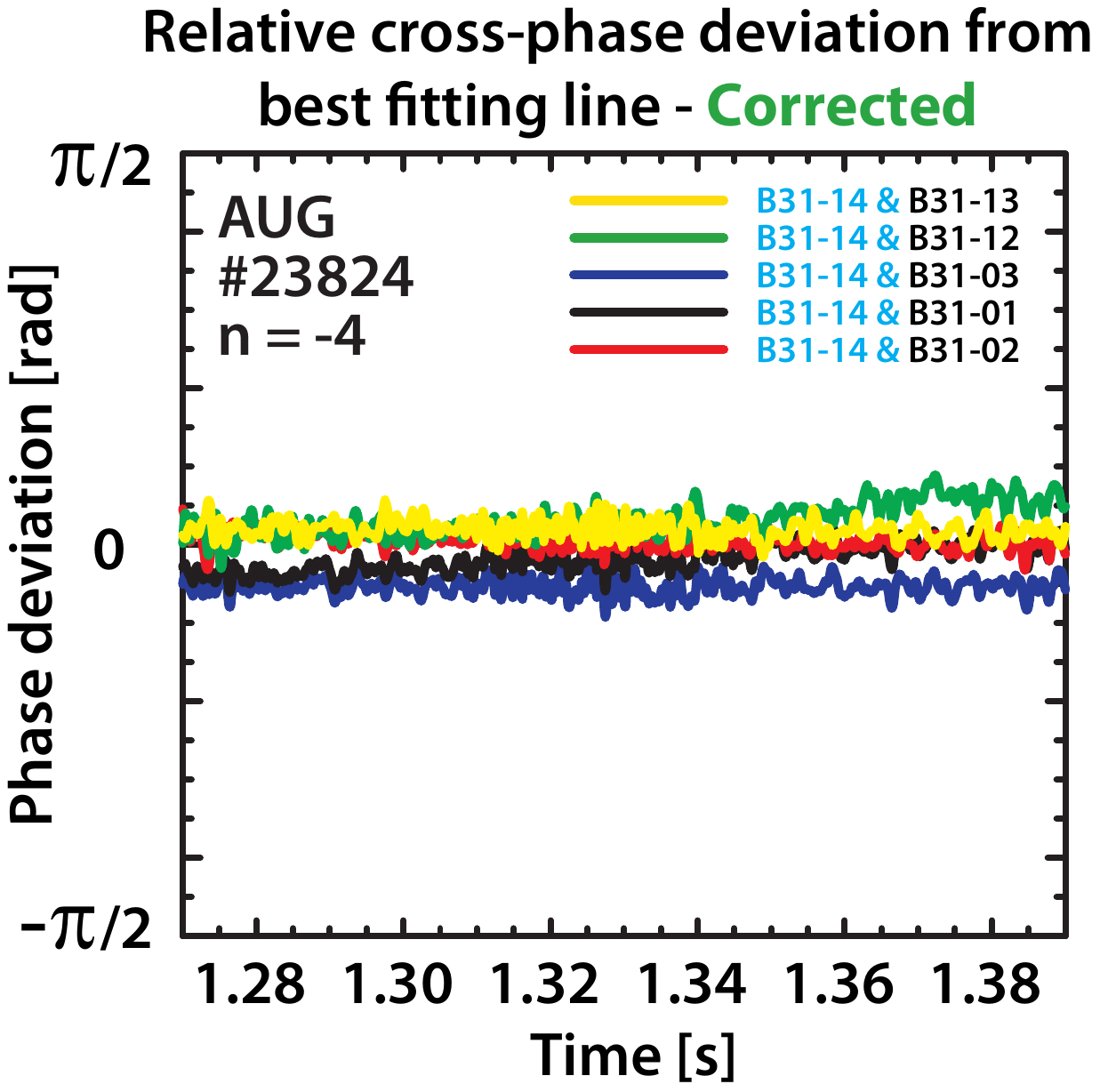}
  \caption{The deviation of the relative phases from the $n = -4$ mode for independent channel pairs as a function of time. The cross-phase correction leads to a significantly better fit compared to the one of figure~\ref{fig:deviation_no}.}
  \label{fig:deviation_corr}
\end{figure}

The deviation of the relative phases from the $n = -4$ for independent channel pairs as a function of time is visible in figure~\ref{fig:deviation_corr}.
Contrary to the results shown in figure~\ref{fig:phasediagram_no}b and figure~\ref{fig:deviation_no}, here $n=-4$  is the best fitting value even in the $[-100, 100]$ interval.
Without correction, the residual of the best fit ($n=-4$) in the interval $[-20, 20]$ is only 2-3 times lower than the residual of close integer numbers such as $-10$, $-16$ and $8$.
Using the correction, the residual of the best fit is lower by more than an order of magnitude than the second best one.

The same comparison with and without cross-phase correction has been performed for different AUG discharges.
In total 11 cases are presented from the period 2006--2013.
The results are presented in table~\ref{tab:qs}.
In most cases an order of magnitude improvement is observed in the residual (defined in \eqref{eq:residual}) of the fits.
However, the remaining deviation still has a systematic component.
Considering the 11 cases presented in this paper, the remaining systematic error has no clear, direct dependence on the frequency or on the toroidal mode number.
We have to note that on AUG there is a slight difference ($\sim 0.01$ rad) in the poloidal position of the toroidal array of balloning coils which - taking the poloidal mode numbers into consideration -- contributes to the remaining systematic error.
Some systematic errors may arise due to the angular orientation of the pick-up coils or due to unknown inaccuracy in the coil position.
Eddy currents in the vessel structures can cause major systematic errors.
Their exact behaviour could only be taken into consideration by simulations~\cite{schittenhelm97analysis}.
\begin{table}[htb!]
\footnotesize
  \begin{center}
    \begin{tabular}{|c|c|c|c||c|c|}
\hline
Shot	&Time range	&Frequency	&Mode	&Average Q w/o	&Average Q with	\\
number	&[s]		&range [kHz]	&number	&correction [rad$^2$]	&correction [rad$^2$]	\\
\hline\hline
\#21030	&1.68-1.94	&160-180	&-4	&5.39	&0.36	\\
\#21030	&1.79-1.94	&170-190	&-5	&4.53	&0.33	\\
\hline
\#23824	&1.27-1.39	&130-160	&-4	&7.91	&0.29	\\
\#23824	&1.22-1.30	&115-135	&-3	&9.17	&0.97	\\
\#23824	&1.15-1.19	&115-125	&-4	&8.62	&0.23	\\
\#23824	&1.30-1.45	&140-180	&-5	&6.46	&0.39	\\
\hline
\#26611	&1.60-1.73	&160-200	&-3	&3.97	&1.44	\\
\#26611	&1.60-1.73	&170-210	&-4	&10.71	&7.83	\\
\#26611	&1.63-1.73	&180-210	&-5	&4.39	&0.65	\\
\hline
\#29586	&1.16-1.24	&180-220	&-3	&2.27	&0.51	\\
\#29586	&1.16-1.24	&230-190	&-4	&1.33	&0.83	\\
\hline
    \end{tabular}
    \caption{\label{tab:qs} The average residuals ($Q$) defined in \eqref{eq:residual} of mode number fits for the investigated shots. In most cases an order of magnitude improvement is achieved by the correction of cross-phases.}
  \end{center}
\end{table}

\begin{figure}[htb!]\centering
  \includegraphics[width = 80mm]{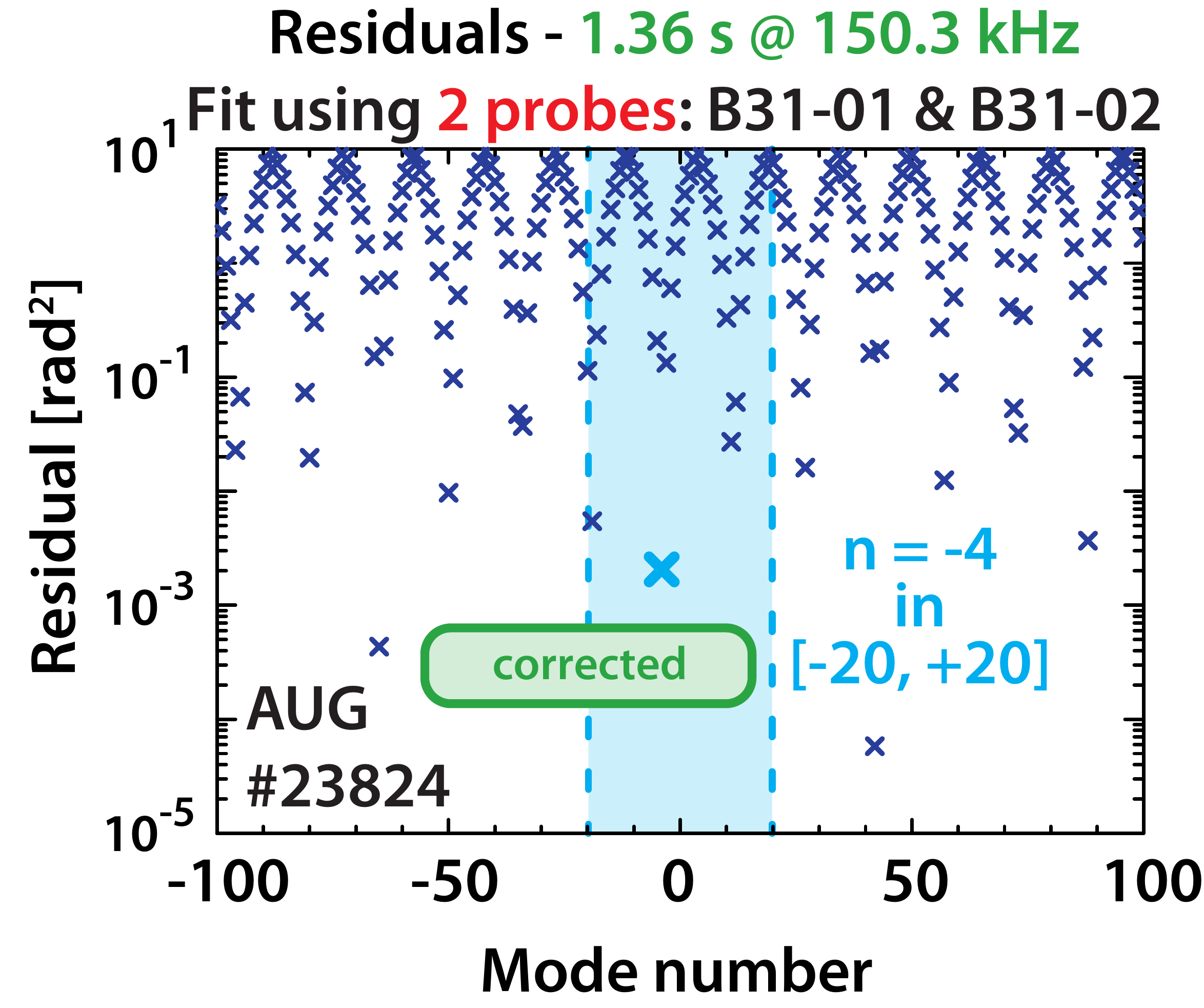}
  \caption{The residuals defined in \eqref{eq:residual} are shown as a function of the mode number when only two probes (namely B31-01 and B31-02) were used to characterize the mode structure. With accurate measurements, fewer detectors are sufficient to provide adequate mode number determination. The periodicity of the residuals is because the probe placing has a near fifteen fold symmetry ($\phi/\pi = 0.131 \sim 2/15$).}
  \label{fig:two_probes}
\end{figure}

An important corollary of the systematic error elimination is that one can investigate mode numbers with fewer detectors.
This has been demonstrated by choosing two coils with a relative probe position $\phi/\pi$ = 0.131, so that $\phi/\pi$ is not a low order rational.
The residual of the linear fit as a function of mode number is shown in figure~\ref{fig:two_probes}.
It shows that even two well positioned probes can provide accurate mode numbers if the magnetic field fluctuation measurement is precise.
However, since $\phi/\pi$ is close to $2/15 \sim 0.133$ the residual of fits with mode numbers smaller or greater than $−4$ by the multiples of 15 are also small as is visible in figure~\ref{fig:two_probes}.

The use of fewer probes can be beneficial in cases when the number of available coils is limited.
This could happen, for example, when some of the signals are saturated.
The spacing of the two coils affects the range of mode numbers which can be observed.
As expressed by Hole et. al. \cite{hole09high}, large coil spacing provides fine resolution, but in order to identify high mode numbers small coil spacing is needed.
It is important to note that a spatial aliasing could arise due to the symmetries in the coil spacing.
Similarly to the case of digitizing time signals, a kind of analogue filter is needed to eliminate the aliasing effect.
A cut-off arising from the coil dimensions can serve as an analogue filter, since it cannot detect modes with smaller wavelength than its size.
For a high, but not unreasonable toroidal mode number ($n \approx 50$), the coil dimensions would be unacceptable.
Thus, in a complex set-up with several probes, at least two coils with small spacing are suggested in order to decide whether mode number is real or an artefact of spatial aliasing.

\section{Conclusion}\label{sec:conc}

In this paper we describe the effect of the magnetic coil transfer functions on mode number analysis.
The correct characterization of the coil phase transfer functions requires in-situ measurements, since this is the only straightforward way to account for all the effects in the measurement set-up used in the experiments.
A new evaluation of previously recorded in-situ coil response measurement data~\cite{schmid05charakterisierung} on ASDEX Upgrade has shown that the transfer functions of the magnetic pick-up coils differ significantly in the Alfv\'{e}nic frequency range.
Determining the absolute phase shift between the driving perturbation and the pick-up coils is difficult, but for mode number analysis only the relative phase differences are important.
The phases of the relative transfer functions were calculated, and we found that the difference in the relative phase delay can reach up to 1 radian $(60^\circ)$ in the 50-250 kHz frequency range.
It was demonstrated that applying the correction of the transfer function in the mode number analysis leads to considerably better mode number fitting.
With this reduction of systematic errors, mode numbers can be accurately determined using fewer detectors.
The correction method was tested on the toroidal mode number of strong, globally coherent Toroidicity-induced Alfv\'{e}n Eigenmodes.
Eleven example cases are presented from the period 2006--2013.
In most cases an order of magnitude improvement is observed in the residual of the mode number fits.

\section*{Acknowledgments}

The authors are grateful to T. Todd, H. Meyer and G. P\'{o}r for fruitful discussions. Authors at BME NTI acknowledge the support of Hungarian State grant NTP-TDK-14-0022.
This work has been carried out within the framework of the EUROfusion Consortium and has received funding from the Euratom research and training programme 2014-2018 under grant agreement No 633053.
The views and opinions expressed herein do not necessarily reflect those of the European Commission. 

\section*{References}
\addcontentsline{toc}{section}{References}
\bibliographystyle{unsrt}
\bibliography{references}

\begin{thebibliography}{10}

\bibitem{nardone92multichannel}
C.~Nardone.
\newblock Multichannel fluctuation data analysis by the singular value
  decomposition method. {A}pplication to {MHD} modes in {JET}.
\newblock {\em Plasma Physics and Controlled Fusion}, 34(9):1447, 1992.
\newblock \url{http://stacks.iop.org/0741-3335/34/i=9/a=001}.

\bibitem{kim99mhd}
J.~S. Kim, D.~H. Edgell, J.~M. Greene, E.~J. Strait, and M.~S. Chance.
\newblock {MHD} mode identification of tokamak plasmas from {Mirnov} signals.
\newblock {\em Plasma Physics and Controlled Fusion}, 41(11):1399, 1999.
\newblock \url{http://stacks.iop.org/0741-3335/41/i=11/a=307}.

\bibitem{zegenhagen06analysis}
S.~Zegenhagen, A.~Werner, A.~Weller, and T.~Klinger.
\newblock Analysis of {Alfv\'{e}n} eigenmodes in stellarators using non-evenly
  spaced probes.
\newblock {\em Plasma Physics and Controlled Fusion}, 48(9):1333, 2006.
\newblock \url{http://stacks.iop.org/0741-3335/48/i=9/a=005}.

\bibitem{pokol08experimental}
G.~Pokol, G.~Papp, G.~Por, S.~Zoletnik, A.~Weller, and the W7-AS~team.
\newblock Experimental study and simulation of {W7-AS} transient {MHD} modes.
\newblock {\em AIP Conference Proceedings}, 993(1):215--218, 2008.
\newblock
  \url{http://scitation.aip.org/content/aip/proceeding/aipcp/10.1063/1.2909112}.

\bibitem{pokol10wavelet}
G.I. Pokol, N.~Laz\'{a}nyi, G.~P\'{o}r, A.~Magyarkuti, G.~Papp, A.~Gude,
  V.~Igochine, M.~Maraschek, and the ASDEX Upgrade~Team.
\newblock A wavelet based method for detecting transient plasma waves and
  determining their spatial structure.
\newblock In {\em Proceedings of the 37th EPS Conference on Plasma Physics,
  Dublin, 2010}, volume 34A of {\em Europhysics Conference Abstracts}, page
  P5.129, 2010.
\newblock \url{http://ocs.ciemat.es/EPS2010PAP/pdf/P5.129.pdf}.

\bibitem{schittenhelm97analysis}
M.~Schittenhelm, H.~Zohm, and the ASDEX Upgrade~Team.
\newblock Analysis of coupled {MHD} modes with {M}irnov probes in {ASDEX}
  {U}pgrade.
\newblock {\em Nuclear Fusion}, 37(9):1255, 1997.
\newblock \url{http://stacks.iop.org/0029-5515/37/i=9/a=I06}.

\bibitem{igochine03investigation}
V.~Igochine, S.~G\"{u}nter, M.~Maraschek, and the ASDEX Upgrade~Team.
\newblock Investigation of complex {MHD} activity by a combined use of various
  diagnostics.
\newblock {\em Nuclear Fusion}, 43(12):1801, 2003.
\newblock \url{http://stacks.iop.org/0029-5515/43/i=12/a=023}.

\bibitem{heeter00fast}
R.~F. Heeter, A.~F. Fasoli, S.~Ali-Arshad, and J.~M. Moret.
\newblock {F}ast magnetic fluctuation diagnostics for {A}lfv\'{e}n eigenmode
  and magnetohydrodynamics studies at the {J}oint {E}uropean {T}orus.
\newblock {\em Review of Scientific Instruments}, 71(11):4092--4106, 2000.
\newblock \url{http://dx.doi.org/10.1063/1.1313797}.

\bibitem{appel05calibration}
L.~C. Appel and M.~J. Hole.
\newblock {C}alibration of the high-frequency magnetic fluctuation diagnostic
  in plasma devices.
\newblock {\em Review of Scientific Instruments}, 76(9):093505, 2005.
\newblock \url{http://dx.doi.org/10.1063/1.2009107}.

\bibitem{moret98magnetic}
J.-M. Moret, F.~Buhlmann, D.~Fasel, F.~Hofmann, and G.~Tonetti.
\newblock {M}agnetic measurements on the {TCV} tokamak.
\newblock {\em Review of Scientific Instruments}, 69(6):2333--2348, 1998.
\newblock \url{http://dx.doi.org/10.1063/1.1148940}.

\bibitem{reimerdes01mhd}
Holger Reimerdes.
\newblock {\em {MHD} stability limits in the {TCV} tokamak}.
\newblock PhD thesis, \'{E}cole {P}olytechnique {F}\'{e}d\'{e}rale de Lausanne,
  2001.
\newblock \url{http://dx.doi.org/10.5075/epfl-thesis-2399}.

\bibitem{haskey13multichannel}
S.~R. Haskey, B.~D. Blackwell, B.~Seiwald, M.~J. Hole, D.~G. Pretty, J.~Howard,
  and J.~Wach.
\newblock A multichannel magnetic probe system for analysing magnetic
  fluctuations in helical axis plasmas.
\newblock {\em Review of Scientific Instruments}, 84(9):093501, 2013.
\newblock
  \url{http://scitation.aip.org/content/aip/journal/rsi/84/9/10.1063/1.4819250}.

\bibitem{fredrickson95resutls}
E.~Fredrickson, K.~M. McGuire, and Z.~Y. Chang.
\newblock {Results from the TFTR high-frequency toroidal Mirnov array for
  detection of modes in the Alfv\'{e}n range of frequencies}.
\newblock {\em Review of Scientific Instruments}, 66(1):813--815, 1995.
\newblock \url{http://dx.doi.org/10.1063/1.1146529}.

\bibitem{schmid05charakterisierung}
Andreas Schmid.
\newblock Charakterisierung von {P}lasmafluktuationen mit einer kombinierten
  {M}irnov-{L}angmuir-{S}onde am {T}okamak {ASDEX} {U}pgrade.
\newblock Master's thesis, Technische Universit\"{a}t M\"{u}nchen, 2005.
\newblock \url{http://books.google.de/books?id=yzIMcgAACAAJ}.

\bibitem{mallat08wavelet}
Stephane Mallat.
\newblock {\em A Wavelet Tour of Signal Processing, Third Edition: The Sparse
  Way}.
\newblock Academic Press, 3rd edition, 2008.

\bibitem{sertoli13characterization}
M.~Sertoli, L.~Horv\'{a}th, G.I. Pokol, V.~Igochine, L.~Barrera, and the ASDEX
  Upgrade~Team.
\newblock Characterization of saturated {MHD} instabilities through {2D}
  electron temperature profile reconstruction from {1D} {ECE} measurements.
\newblock {\em Nuclear Fusion}, 53(5):053015, 2013.
\newblock \url{http://stacks.iop.org/0029-5515/53/i=5/a=053015}.

\bibitem{pokol13continuous}
G.~I. Pokol, L.~Horv\'{a}th, N.~Lazanyi, G.~Papp, G.~Por, V.~Igochine, and the
  ASDEX Upgrade~Team.
\newblock Continuous linear time-frequency transforms in the analysis of fusion
  plasma transients.
\newblock In {\em Proceedings of the 40th EPS Plasma Physics Conference, Espoo,
  2013}, volume 37D of {\em Europhysics Conference Abstracts}, page P5.116,
  2013.
\newblock \url{http://ocs.ciemat.es/EPS2013PAP/pdf/P5.116.pdf}.

\bibitem{pokol07application}
G.~Pokol, G.~Por, S.~Zoletnik, and the W7-AS~team.
\newblock Application of a bandpower correlation method to the statistical
  analysis of {MHD} bursts in quiescent {Wendelstein-7} {AS} stellarator
  plasmas.
\newblock {\em Plasma Physics and Controlled Fusion}, 49(9):1391, 2007.
\newblock \url{http://stacks.iop.org/0741-3335/49/i=9/a=003}.

\bibitem{wenninger13nonlinear}
Ronald Wenninger.
\newblock {\em {T}he non-linear evolution of edge localized modes}.
\newblock PhD thesis, Ludwig-Maximilians-Universit{\"a}t, M{\"u}nchen, January
  2013.
\newblock \url{http://nbn-resolving.de/urn:nbn:de:bvb:19-152189}.

\bibitem{papp11low}
G.~Papp, G.~I. Pokol, G.~Por, A.~Magyarkuti, N.~Laz\'{a}nyi, L.~Horv\'{a}th,
  V.~Igochine, M.~Maraschek, and the ASDEX Upgrade~Team.
\newblock Low frequency sawtooth precursor activity in {ASDEX} {U}pgrade.
\newblock {\em Plasma Physics and Controlled Fusion}, 53(6):065007, 2011.
\newblock \url{http://stacks.iop.org/0741-3335/53/i=6/a=065007}.

\bibitem{horvath14changes}
L.~Horv\'{a}th, G.~I. Pokol, G.~Papp, G.~Por, Ph.~W. Lauber, V.~Igochine
  A.~Gude, and the ASDEX Upgrade~Team.
\newblock Changes in the radial structure of {EPMs} during the chirping phase
  taking the uncertainties of the time-frequency transforms into account.
\newblock In {\em Proceedings of the 41th EPS Plasma Physics Conference,
  Berlin, 2014}, volume 38F of {\em Europhysics Conference Abstracts}, page
  P2.008, 2014.
\newblock \url{http://ocs.ciemat.es/EPS2014PAP/pdf/P2.008.pdf}.

\bibitem{garciamunoz11fast}
M.~Garcia-Munoz, I.G.J. Classen, B.~Geiger, W.W. Heidbrink, M.A.~Van Zeeland,
  S.~\"{A}k\"{a}slompolo, R.~Bilato, V.~Bobkov, M.~Brambilla, G.D. Conway,
  S.~da~Gra\c{c}a, V.~Igochine, Ph. Lauber, N.~Luhmann, M.~Maraschek, F.~Meo,
  H.~Park, M.~Schneller, G.~Tardini, and the ASDEX Upgrade~Team.
\newblock {F}ast-ion transport induced by {A}lfv\'{e}n eigenmodes in the
  {ASDEX} {U}pgrade tokamak.
\newblock {\em Nuclear Fusion}, 51(10):103013, 2011.
\newblock \url{http://stacks.iop.org/0029-5515/51/i=10/a=103013}.

\bibitem{hole09high}
M.~J. Hole, L.~C. Appel, and R.~Martin.
\newblock {A high resolution Mirnov array for the Mega Ampere Spherical
  Tokamak}.
\newblock {\em Review of Scientific Instruments}, 80(12):123507, 2009.
\newblock \url{http://dx.doi.org/10.1063/1.3272713}.

\end{thebibliography}

\end{document}